# Extended topological defects as sources and outlets of dislocations in spherical hexagonal crystals

D. S. Roshal, K.Yu. Petrov, A.E. Myasnikova, and S.B. Rochal

Faculty of Physics, Southern Federal University, 5 Zorge str., 344090 Rostov-on-Don, Russia

Extended topological defects (ETDs) arising in spherical hexagonal crystals due to their curvature are considered. These prevalent defects carry a unit total topological charge and are surrounded by scalene pentagonal boundaries. Topological peculiarities of reactions between ETDs and dislocations are considered. Similarly to boundaries of the usual planar crystalline order the ETDs emit and absorb the dislocations without preservation of their dislocational charge. Dislocations located inside the ETD area lose it and the enforced ETD decay can proceed in different ways without conservation of the total Burgers vector of the dislocations emitted.

Two-dimensional (2D) ordered structures with an unusual topology are under discussion since the very beginning of the 20$^{th}$ century. Trying to explain the periodic law of Mendeleev, J. J. Thomson proposed a model of atom, according to which the electrons confined at the sphere surface interact by means of Coulomb potential. Determination of the equilibrium position of repelling equally-charged particles on the sphere was called the Thomson problem [1]. Later, it was generalized to the case of non-Coulomb potentials [2]. Tammes considered the similar problem how *N* identical spherical caps should be packed on the sphere to provide the maximal cap size [3].

Experimental investigation of the behavior of colloidal particles located at the interface between two liquids was started by Ramsden [4] in 1903. More than a century later this study led to the synthesis of nanoporous capsules - colloidosoms [5]. Similar ordered structures appear in various systems. For example, they are formed by viral capsid proteins [6, 7], localized electrons in multi-electron bubbles in superfluid helium [8], Pickering emulsion on spherical surfaces [9, 10, 11] and even occur in coding theory [12, 13]. All these natural and synthetic objects are more or less ordered structures forming the 2D closed shells topologically equivalent to a sphere.

Due to the curved topology new crystallographic peculiarities appear in these systems [14]. One of such peculiarities is the inevitable existence of topological defects that causes the curvature of the ordered 2D spherical structures. Depending on presence of the other types of defects the spherical structures can be divided into two groups. The viral capsids from the first group demonstrate the 'perfect' spherical crystalline [6] or quasicrystalline [15] structures with the regular curvature-related topological defects only. The spherical crystals from the second group are more disordered. Examples are solid colloidosoms and 2D colloidal crystals formed on the spherical surfaces [5, 9, 11]. The important features of these systems are presence of dislocations usual for the planar hexagonal lattice and not so symmetric arrangement of topological defects, which often take the form of scars [10]. The other 'exotic' ETDs unconventional for the planar geometry are also possible. The related defect motifs were studied in the frame of the Thomson problem [16,17]. Recently, formation of ETDs with a square order inside on the colloidosome surface [5], was explained [18] in the frame of Lennard-Jones inter-particle coupling.

The pioneer experimental work [10] devoted to the peculiarities of the spherical order in colloidal crystals was published a decade ago. It was found that this order is very sensitive to the ratio $R/a$, where $R$ is the radius of the sphere and $a$ is an average particle radius. For $R/a \geq 5$ the authors of Ref. [10] have found linear defects, which they called grain boundaries, or scars. These defects present the chains consisting of closely located particles with different surroundings. Particles having 5 or 7 nearest neighbors sequentially alternate in the scars, while the other particles have 6 neighbors. Therefore the scar is usually treated as a sequence of elementary 5-fold and 7-fold disclinations. These defects have been investigated in subsequent experimental and theoretical studies [11,19]. The theory [19] successfully considers the spherical hexagonal order in colloidal crystals as a result of simple repulsion of particles retained on a spherical surface and explains formation of scars in the frames of continuum approach. However, as demonstrated below for the case of the simplest repulsive pair potentials, the emerging spherical hexagonal order is essentially distorted in the relatively wide areas of ETDs, where only the Delaunay triangulation [20] can match the nodes with their neighbors and thus localize the relatively narrow scars.

The aim of this Letter is to demonstrate new topological properties of ETDs arising in spherical crystals due to their curvature. Here we show that: i) enforced decay of the ETD does not conserve the total Burgers vector of the dislocations emitted; ii) dislocations within the ETD area lose their dislocational charge and the order outside the defect doesn't display their existence in any way.

We start our consideration from some more or less known peculiarities of spherical hexagonal order. Let us recall that the self-assembly of 2D structure on a non-planar surface can be described by the conditional minimization of the system free energy $F$ with respect to coordinates of the system particles. The condition imposed is that any particle during minimization should be on the surface under consideration. The order in the spherical colloidal crystals is successfully modeled and analytically studied in the frame of the simplest power low pair potentials [16,17,19] when the free energy has the form:

$$F = \varepsilon \sum_{j>i}^{N} \frac{1}{r_{ij}^{\alpha}}, \qquad (1)$$

where $r_{ij}$ is the distance between $i^{th}$ and $j^{th}$ particles, $N$ is the number of particles. The exponent $\alpha = 1$ for Coulombic long range interaction of charges, while the solution of Tammes problem [3] (very short-range interaction) corresponds to $\alpha \to \infty$. The interaction of particles by means of Lennard-Jones pair potential is also reduced to energy (1) with $\alpha = 12$ provided the particles on a sphere are located closely enough and the term associated with their repulsion prevails over their attraction. Note also that the conditional minimization of Eq. (1) yields different equilibrium structures corresponding to the same values of $N$ and $\alpha$ depending on the initial distribution of particles.

Numerically obtained spherical structure with $N = 700$ particle and $\alpha = 12$ (see Fig. 1) is a typical one for the $N$ range from 400 to 1000 and for the $\alpha$ values in the range of several tens (for larger $\alpha$ the numerical minimization of (1) becomes difficult). Analogous structures corresponding to minima of energy (1) can be also obtained with the help of public domain programs, see for example [21]. Global-minima spherical structures with $\alpha = 1$ are extensively studied. The putative list of them for N<20000 can be found at the same site, see also [17]. The hexagonal order in the global-minima structures is more perfect than in the local-minima ones and the areas occupied by the extended defects are smaller. However, for the same $N$ value the equilibrium energies of global and local-minima structures are very close [14]. If this value increases, the difference between the equilibrium energies is reduced and the number of hexagonal structures with the similar energies, but different arrangement of particles in defects, grows exponentially. [22]. Due to this fact, it is much more probable to observe experimentally a colloidal crystal corresponding to one of numerous local minima than that with the globally minimal free energy. This point of view is also supported by our observation that ETDs in experimental colloidal crystals [5,10,11] are more complicated (for example, the scars are longer) than the defects in the global-minima theoretical spherical structures presented in [21].

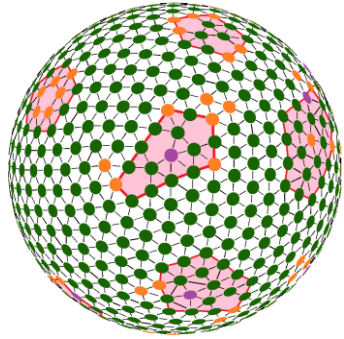

**Fig. 1** (Color online). *Spherical structure with N=700 particles. Extended topological defects with the unit positive topological charge are highlighted by red pentagons. Orange and purple circles denote particles with the smallest ($E_p<0.5\ E_{avg}$) and the largest ($E_p>1.34\ E_{avg}$) energy per particle, respectively. Here $E_p$ is the energy per particle and $E_{avg}$ is the average energy per particle.*

Note that the total topological charge [10] of the ETD is completely determined by the number of sides of characteristic polygon, surrounding the defect provided the polygon satisfies two following conditions: 1) the polygon sides pass only through the nodes with six neighbors; 2) the angle between the nearest polygon sides in the initial planar hexagonal order is equal to $2\pi/3$. Then the total topological charge $q$ of the defect is simply defined as

$$q = 6 - m, \qquad (2)$$

where $m$ is the number of sides of the characteristic polygon. The conventional dislocations without a topological charge can always be surrounded by hexagons. The ETDs with the total charge $q=1$ are surrounded by pentagons which are scalene in general case. Let us recall that the simplest local 5-fold disclination (surrounded by the regular pentagon) is usually associated with elimination of the $\pi/3$ sector from the hexagonal planar lattice [14]. In general, the edges of the eliminated sector can be glued after some relative shift equal to a translation of the initial hexagonal order. Such a shift implies that the characteristic pentagon is scalene and the initial hexagonal order near the top of the resulting solid angle is broken strongly within the defect area. But these peculiarities cannot change the total topological charge of the defect provided it is surrounded by the outer hexagonal order.

We have obtained about 50 spherical structures, with the number of ordered repulsive particles from 700 to 1000. Different initial random distributions of particles and different algorithms of energy (1) minimization (including the algorithm [23]) regularly resulted in appearance of extended areas with essentially distorted hexagonal order. In all the cases, the hexagonal order surrounding the defects was global. We have found no defects which do not allow the continuous circulation around. This particle arrangement corresponds always to more or less defective mapping of a single planar hexagonal lattice onto the sphere by means of the icosahedron net. We were always able to localize exactly twelve ETDs surrounded by pentagons. These defects repel each other and are located approximately near the vertices of an icosahedron (see fig. 1). Spherical structures with arrangement of defects near the vertices of an icosahedron were obtained theoretically [19] and observed experimentally [11]. These topologically induced defects of spherical hexagonal order are usually treated as linear scars [10,11,19]. However, as we explain below such an interpretation of the defects is incomplete.

Let us recall, that the scars were initially defined as 'high-angle (30$^o$) grain boundaries, which terminate freely within the crystal' [10]. Later it was understood that the scars have a variable rotation angle from 0$^o$ (at the scar ends) to 30$^o$ (at the scar center) [11]. Our numerical simulations demonstrate that inside the ETD area the structure is strongly disordered (see figures 1 and 2) and without the formal triangulation it is impossible to relate the nodes with their neighbors. Therefore, it is correct to speak about the hexagonal order and to discuss the local order orientation only in the region outside the ETD area. This statement is justified by an analysis of energy per one particle of the spherical structures. Particles with the maximal deviations of this energy from the average value are located in the vicinity of ETD areas (see fig. 1) *but these 'anomalous' particles do not form the liner chains like scars*. The Delaunay triangulation assigns 6 neighbors to the nodes, which are not related obviously to the hexagonal order (see fig. 2) and thus it localizes the relatively narrow scar being only a part of the extended defect area. In our opinion, any conceivable analysis based purely on the particle energy cannot detect the scars. In addition, let us stress that we use our own software only to demonstrate that the initial regular hexagonal order is actually absent in a relatively wide defect region around the scar. The presentation of spherical structures in the form chosen in the program [23] obscures this peculiarity.

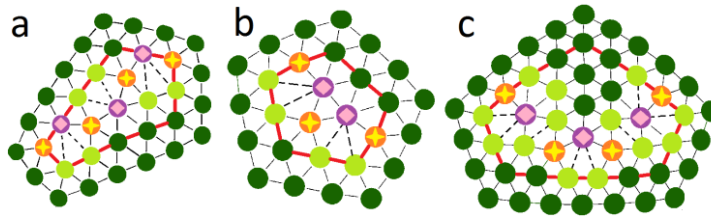

**Fig. 2** (Color online). *Triangulation of the order inside the extended defect region creates the scars. Three extended defects are shown. The nodes with 5 and 7 neighbors form the scars and they are colored in orange and violet. The light-green nodes according to Delaunay triangulation have six neighbors but do not belong to the hexagonal order. If these nodes are taken into account the defects take rather extended than linear form. Panels (a-b) present two ETDs of the spherical structure shown in fig 5(a), while the panel (c) demonstrates the ETD of the other spherical structure formed by 1000 particles.*

The fact that the scars present the sequences of elementary 5-fold and 7-fold disclinations suggests the possibility to consider the scar as a chain of associated dislocations [10,19]. Indeed, it is easy to verify that the simplest dislocation of the planar hexagonal order can be represented by a pair of neighboring elementary 5-fold and 7-fold disclinations. The corresponding dislocation has Burgers vector equal to the minimal translation of the hexagonal order. Using a virtual optical tweezers (see a similar real experiment in [24]) we induce two reactions (see fig. 3) of dislocation emission by the ETD shown in fig. 2 (a). These enforced reactions are interesting because in both cases the elementary 5-fold disclination is detached and it carries out all the topological charge of the ETD.

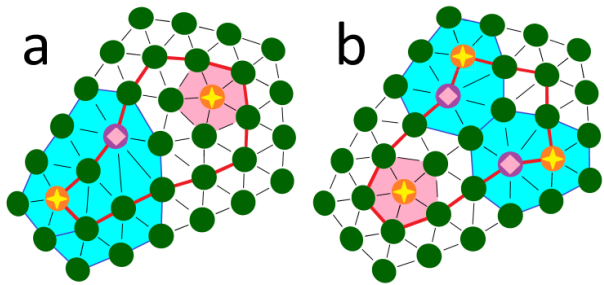

**Fig**. 3 (Color online). *Two possible ways to decompose the ETD (located initially within the red boundary in fig. 2(a)) by the application of external forces. Both reactions are mainly associated with the order reconstruction inside the boundary. Four inner positions are reorganized to form a rhombus (a) or a linear chain (b). In both enforced reactions the elementary 5-fold disclination is detached and it carries out all the topological charge of the ETD. The discination region is colored in red. Hexagons containing the dislocations are colored in blue. The length of the Burgers vector of the dislocation (a) is equal to 2, while the length of the total Burgers vector of two dislocations shown in panel (b) is equal to $\sqrt{3}$.*

Comparison of fig 2 (a) and fig. 3 demonstrates clearly that in course of the ETD decomposition the elementary disclinations with the opposite topological charges can annihilate in pairs.. Moreover, fig. 2 shows that the total Burgers vector is not conserved in reactions with participation of localized disclinations. Let as clarify this not obvious fact. At the first glance (see fig. 4) it seems that if the characteristic pentagon is scalene then an excess line of nodes crosses its longer side and ends with the dislocation inside the pentagon, where a dislocational charge is localized. However, the last statement is incorrect and an analysis of the order outside the defect area proves this fact.

Let us note that the characteristic pentagon can be changed by adding strips, parallel to its sides (see Fig. 4). Thus, after adding a strip the pentagon side parallel to the strip becomes one node shorter, and each of two adjacent sides becomes one node longer. As a result, the contour perimeter is increased by one node. Let us apply the adding strip operations to the scalene characteristic pentagon, which is shown in red in Fig. 4. Adding four strips as it is demonstrated in Fig. 4 makes the characteristic pentagon equilateral. It means that the outside order is perfect and the distorted structure in the initial scalene characteristic pentagon does not possess a dislocational charge.

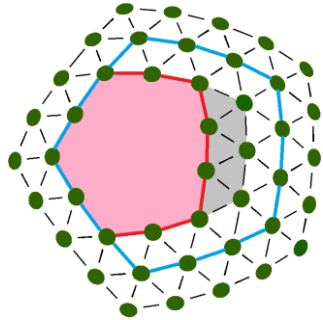

**Fig. 4** (Color online). *The hexagonal order outside the scalene characteristic pentagon (shown in red). It seems that an excess line of nodes crosses its longer side and ends with the dislocation inside the defect area. However, characteristic pentagon can be changed by the adding strips, parallel to its sides. One additional strip (shown in grey) converts the pentagon with the sides <2,2,3,2,2> into the pentagon <2,3,2,3,2>. Adding of four strips transforms the initial scalene pentagon into the equilateral one <3,3,3,3,3> (shown in blue). The order outside the outer equilateral pentagon is obviously perfect and no sign of dislocational charge can be found.*

Let us study the problem for an arbitrary scalene characteristic pentagon, which can be characterized by a five-dimensional integer vector $\mathbf{S}=<n_1, n_2, n_3, n_4, n_5>$, its components are equal to the lengths of the pentagon sides. Adding a strip parallel to the first side of the contour is equivalent to adding the translation

$$\mathbf{a_1} =<-1, 1, 0, 0, 1> \tag{3}$$

to the vector $\mathbf{S}$. The adding operations for four other sides correspond to translations obtained from vector (3) by the cyclic permutation of its components. Let the symmetric matrix $M_{ij}$ consists of lines $\mathbf{a_i}$. Then

$$\Delta S_i = M_{ij} V_j, \quad (4)$$

where the component $\Delta S_i$ is equal to a length change of *i*-th side, and component $V_j$ specifies how many strips were added to *j*-th side.

Matrix $M_{ij}^{-1}$ entering the opposite relationship

$$V_i = M_{ij}^{-1} \Delta S_j \quad (5)$$

is integer:

$$M_{ij}^{-1} = \begin{bmatrix} -1 & 0 & 1 & 1 & 0 \\ 0 & -1 & 0 & 1 & 1 \\ 1 & 0 & -1 & 0 & 1 \\ 1 & 1 & 0 & -1 & 0 \\ 0 & 1 & 1 & 0 & -1 \end{bmatrix} \quad (6)$$

The rather surprising fact that all coefficients of matrix (6) are integer means that any pentagon surrounding the topological defect can be transformed into the equilateral pentagon by adding strips. Using the elaborated above mathematics one can say that applying a set of operations **V**=<0,1,2,1, 0> to the scalene pentagon **S**=<2,2,3,2,2> shown in Fig. 4 one can transform it into the equilateral pentagon with the side length equal to three (see Fig. 4). From the physical point of view the integer form of matrix (6) means that absorbed dislocations are integrated into the ETD structure completely, since the order outside the defect does not display their existence and these dislocations have lost the dislocational charge. This effect is similar to the output of dislocations to the crystalline boundary, where they are eliminated completely.

Note also that the ETDs with unit negative topological charge (surrounded by heptagons) have similar topological properties and also absorb the dislocations completely. The matrix (analogous to the matrix (6)) describing the variation of heptagon sides consists of integer coefficients and a similar consideration proves that the ETD with $q=-1$ cannot possess a dislocational charge. However, to produce this defect one should insert the π/3 sectors into the planar hexagonal order, which creates an extended area with negative Gaussian curvature near the top of the added sector. Thus this ETD is incompatible with the spherical geometry and its appearance can be expected on the more complex surfaces, in regions where the surface has a saddle shape. See the recent experimental study of colloidal crystals on the surface with the negative Gaussian curvature [25] and the theoretical work [26] devoted to the defect motifs induced by this type of curvature.

The above mathematics provides a new classification of spherical structures with hexagonal order and of ETDs arising in them. In the structures under consideration any scalene characteristic pentagon with the ETD inside can be transformed by adding strips into the equilateral pentagon. After that a perfect hexagonal order with a single central 5-fold disclination can be restored inside the equilateral pentagon. This restoration requires inclusion of $N_v$ additional nodes. Therefore *any ETD with q=1 can be characterized by the invariant number $N_v$ of particles, which should be excluded from the vicinity of the local 5-fold disclination to construct this ETD*. Fig. 5(a) demonstrates how to restore an ideal order for the ETD, which was previously shown in Fig. 2(a). In the ideal case the scalene pentagon <1,4,2,2,4> (see the same contour in Fig. 2(a)) contains inside its boundary ten positions. In the real defect (see Fig. 2(a)) only four of them are filled. So, $N_v = 10-4=6$.

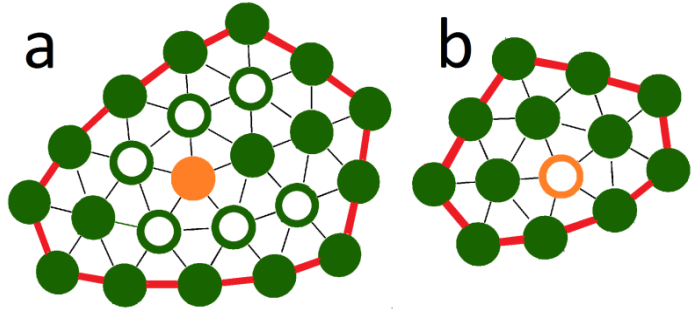

**Fig. 5** (Color online). *Reconstruction of the hexagonal order within the defect area allows classifying the defects. Panel (a) demonstrates that the ETD shown in Fig. 2(a) can be considered as a single 5-fold disclination with six associated vacancies, while the panel (b) shows the simplest frequent defect, which can be treated as superposition of one disclination and one vacancy. After the triangulation this defect looks as the scar presenting a sequence of 5-fold, 7-fold and 5-fold disclinations (see similar scars in fig. 7 of Ref. 19).*

Our idea to restore the ideal order with a single central 5-fold disclination in the defect area develops the approach [17, 27], which is inverse to our one in a manner. In the works [17,18] several low-energy spherical structures with the global $I$ and $I_h$ symmetry were obtained as a result of elimination of some nodes from the initial perfect order corresponding to the spherical regular icosahedron [6] with $10T+2$ positions, where T is the triangulation number: $T=h^2+k^2+hk$; $h$ and $k$ are integer. Our results allow constructing a similar parent *irregular* icosahedron for any low-symmetry spherical structure possessing the global hexagonal order with twelve *ETDs each of them is characterized by the unit topological charge*. Like the net of regular icosahedron the net of irregular one consists of 20 triangles, which are scalene in general case. The edges of these triangles are also translations of the hexagonal order. However the lengths of these translations are equal only approximately and the irregular icosahedron cannot be characterized by a single triangulation number T. See some examples of irregular icosahedron nets in [28]. After the restoration of the perfect order in the areas of all ETDs, the twelve isolated disclinations obtained are matched with the vertices of the irregular spherical icosahedron and its planar net can be constructed. Thus any spherical structure under consideration can be characterized by the certain icosahedron (irregular in general case) with the net cut from the perfect planar hexagonal lattice. Of course, to return from this icosahedron to the initial spherical structure, $N_v(i)$ nodes ($i=1, 2…12$) should be excluded in vicinities of twelve icosahedon vertices. Possibly, the construction of slightly irregular icosahedrons with the subsequent elimination of some nodes in vicinities of their vertices is useful not only for the classification of spherical structure. May be, similar approach would open a new way to search the lowest-energy structures in the frame of the Thomson problem.

Unfortunately, the classification of ETDs based on the $N_v$ number is not unique since the different defects can be characterized by the same $N_v$ value. However, our numerical simulations and preliminary analysis of lowest-energy spherical structures ($400<N<1000$) presented in literature and internet demonstrate that the ETDs with $N_v \leq 3$ are prevalent and the number of different defects with small $N_v$ values is very limited. The only existing ETD with $N_v=1$ corresponds always to a vacancy coinciding with the centre of isolated pentagonal disclination (see Fig. 5 (b)). In the lowest-energy structures [21] the resulting empty pentagon becomes flattened and this defect is usually interpreted as *a sequence of 5-fold, 7-fold and 5-fold disclinations*. Fig. 2 (b) presents one of the ETDs with $N_v = 3$, which was found in the local-minimum structure composed by N=700 particles and shown in Fig. 1. An example of ETD with $N_v = 2$ is presented in the lowest-energy structure with the same number of particles [21]. This structure has four defects of this type. Each of them is surrounded by pentagonal contour like

<2,3,1,3,2> up to the cyclic permutation. After the Delaunay triangulation the defect looks like a simplest dislocation in the vicinity of a single disclination.

The fact that any ETD can be surrounded by an outer equilateral pentagon allows assuming that application of appropriate external forces could easily rotate the ETD as a whole around its centre by angles multiple to $2\pi/5$. Of course, the possibility to rotate the defect as a whole assumes that the defect structure is sufficiently stable and it is easier to rotate the defect than to rebuild it completely. We studied this problem numerically on example of defect shown in Fig. 2(a). To check the defect stability we shifted analogously to Fig. 5(a) the particles in vicinity of the defect region. After that we have minimized energy (1) and all particles have acquired their initial positions (shown in Fig. 1) again. We also checked that enforced reconstructions (shown in Fig. 3(a-b)) of this ETD are reversible. After turning off the external forces and minimizing energy (1) the ETD acquires its initial form. Then we started several times the energy minimizations after different random shifts of particles forming this ETD. In some cases the ETD has taken its initial structure, but it was rotated around its center as a whole. The rotational angle was always multiple to $2\pi/5$. This result is explained by the fact that the location of the ETD centre inside the characteristic pentagon is completely determined by the vector **S** components. The $2\pi/5$ rotation of the ETD conserves its center and simply permutes the vector **S** components cyclically. Such a rotation requires only the order reconstruction in vicinity of the defect center. In contrast, the ETD shift as a whole is impossible without emission or absorption of dislocations since the pentagonal contour surrounding the ETD is not translationally invariant. Any hexagonal translation breaks the characteristic pentagon and changes at least one of its sides. To compensate this change the node lines should be added or eliminated. Thus to shift the ETD an extensive dislocation motion is required. Therefore, in spherical colloidal crystal it should be much easier to rotate the scar than to decompose or to move it. It is very interesting to check this conclusion experimentally.

In conclusion, this Letter deals with the peculiarities of spherical hexagonal order, which is observed in spherical colloidal crystals and is usually modeled by simplest repulsive pair potentials. Following this assumption we studied the spherical order of $N$ particles ($400<N<1000$). Likely previous works, the hexagonal order simulated was found to be global, and twelve relatively wide regions of ETDs were pushed outside of it. In our opinion, only the triangulation of the distorted order inside these regions detects the scars, which are widely discussed now, and present the linear chains of 5-fold and 7-fold elementary disclinations. In contrast to the previous point, three results of the Letter summarized below are not based on the computer simulations. First, we propose to classify the spherical hexagonal structures and ETDs in them using a virtual restoration of the perfect order inside the ETD area. Second, we show that the dislocations located in the ETD area lose the dislocation charge and the hexagonal order surrounding the defect area does not demonstrate its existence in any way. Similarly to linear boundaries of the planar crystalline order, the ETDs emit and absorb the dislocations without preservation of their dislocational charge. Both processes change the particle arrangement in the defect region and shift its center. Third, we demonstrate that the rotation of the ETD as a whole in contrast to its shift does not involve the dislocation motion. Therefore an application of appropriate external forces could rotate the ETD (or the scar) as a whole by angle multiple to $2\pi/5$. This theoretical prediction is interesting for future experimental tests using the experimental techniques of topological optical tweezers [24].

## Acknowledgements

Authors acknowledge financial support of the RFBR grant 13-02-12085 ofi_m.